\documentclass[twocolumn,showpacs,preprintnumbers,a4paper,prl,notitlepage]{revtex4}
\usepackage{graphicx, float}
\usepackage[caption=false]{subfig}
\usepackage{amssymb, amsmath,amssymb,amsfonts}
\usepackage{dcolumn,lipsum}
\usepackage{bm}
\usepackage{color}
\usepackage[utf8]{inputenc}
\usepackage[english]{babel}
\usepackage{dsfont}

\begin{document}

\title{Hamiltonian control of Kuramoto oscillators}

\author{Oltiana Gjata$^{a,b}$, Malbor Asllani$^b$, Luigi Barletti$^a$, Timoteo Carletti$^b$}
\affiliation{$^a$Department of Mathematics $\&$ Computer Science, University of Florence, Viale Morgagni 67/A, 50134 Florence, Italy}
\affiliation{$^b$naXys, Namur Center for Complex Systems, University of Namur, rempart de la Vierge 8, B 5000 Namur, Belgium}

\begin{abstract}
Many coordination phenomena are based on a synchronisation process, whose global behaviour emerges from the interactions among the individual parts. Often in Nature, such self-organising mechanism allows the system to behave as a whole and thus grounding its very first existence, or expected functioning, on such process. There are however cases where synchronisation acts against the stability of the system; for instance in the case of engineered structures, resonances among sub parts can destabilise the whole system. In this Letter we propose an innovative control method to tackle the synchronisation process based on the use of the Hamiltonian control theory, by adding a small control term to the system we are able to impede the onset of the synchronisation. We present our results on the paradigmatic Kuramoto model but the applicability domain is far more large.
\end{abstract}


\maketitle

Synchronisation is one of the most important example of collective behaviour in Nature, being at the basis of many processes in living beings~\cite{Pikovsky2003,Manrubia2005,Boccaletti2008}. Indeed, their activity is regulated by (almost) periodic processes of different duration that must run at unison to determine a collective behaviour~\cite{strogatz,winfree} enable to sustain life. For this reason synchronisation is widespread and it has been studied in many research domains such as biology (flashing of fireflies \cite{buck} and the cricket chirping \cite{walker} during the mating season), chemistry (glycolytic oscillations in populations of yeast cells \cite{richard}), physics (arrays of coupled lasers \cite{laser} and the superconducting Josephson junctions \cite{joseph}), just to mention few of them. One of the most representative case being the heart~\cite{peskin}, an organ of vital importance for all species in the animal kingdom; the heart is composed by a collection of individual cells, myocytes, whose complex interactions among them are responsible for the ability to pump blood in the circulatory system. Initially, at embryonic state, such cells do not interact each other and their beats are independent, only after a couple of days, the myocytes form interconnected sheets of cells that help them beat in unison. The absence of such synchronisation phenomenon in human induces cardiac arrhythmia and artificial pacemakers are necessary to recover the normal behaviour.

Despite the very different nature of the systems exhibiting synchronisation phenomena, most of the main features are quite universal and can thus be described using the paradigmatic Kuramoto model (KM)~\cite{kuramoto1975,kuramoto,syncopen,acebron2005} of coupled non-linear oscillators. Once the $N$ oscillators are set on top of a complex network~\cite{Arenas2008}, the KM can be described by
\begin{equation}
\dot{\phi}_i = \omega_i + \frac{K}{N}\sum_{j=1}^N A_{ij}\sin(\phi_j-\phi_i)\, ,
\label{eq:kurmodnet}
\end{equation}
where $K$ is the interaction strength, $\omega_i$ are the natural frequencies of the oscillators drawn from some distribution $g(\omega)$ and $A_{ij}$ the undirected network adjacency matrix, i.e. $A_{ij}=A_{ji}=1$ if oscillators $i$ and $j$ are directly coupled and zero otherwise. The factor $1/N$ assures a correct behaviour~\footnote{Let us observe that other choices are possible as described in~\cite{Arenas2008}.} of the model~\eqref{eq:kurmodnet} in the thermodynamic limit $N\rightarrow\infty$. Notice that the original KM correspond to an all-to-all coupling~\cite{kuramoto1975}, i.e. $A_{ij}=1$ for all $i\neq j$ and $A_{ii}=0$.

Let us observe that there are cases where such collective rhythm has a negative impact on the organism life~\cite{notsync}; for instance, it has been observed, that certain psychomotor symptoms, e.g. tremors, are results of an abnormal synchronisation phenomenon in the activity of the responsible neuronal zones, whose outcome is the onset of diseases such as epilepsy or Parkinson~\cite{glass}. These issues related to undesired level of synchronisation are well known to engineerings; consider for instance the case of the Millennium Bridge in London~\cite{millenium}, soon after its inauguration, dangerous lateral vibrations appeared, causing its immediate closure. Almost negligible, horizontal oscillations of the bridge can be amplified by walking pedestrians creating thus a positive feedback resulting in larger oscillations (crowd synchronisation) possibly leading to major damages in the bridge structure.

Based on the above observations, scholars quickly realised the great importance of control~\cite{callier1991,locatelli2001,Hinrichsen2005} and possibly impeding the collective synchronisation to prevent  negative undesired effects. In this Letter we introduce an effective control method for the Kuramoto system, inspired by the Hamiltonian control theory~\cite{vittot, chandre} already successfully applied in other frameworks, such as plasmas fusion~\cite{ciraolo} and particles accelerators~\cite{carletti}. The KM is a dissipative system, however recently Witthaut and Timme proposed~\cite{timme} an Hamiltonian model which embeds the Kuramoto model, roughly speaking the system possesses an invariant torus upon which the Hamiltonian dynamics is the same of the original KM. Moreover it can be shown that the Kuramoto oscillators are phase-locked if and only if the Hamiltonian invariant torus is unstable. Hence controlling the Hamiltonian systems to achieve, or not, the stability of the invariant torus, results to be a suitable strategy to control the synchronisation phenomenon in the Kuramoto model~\cite{syncopen}.

Consider the following $N$ dimensional Hamiltonian system, which generalises the one proposed in~\cite{timme} to the case of a network of $N$ oscillators
\begin{eqnarray}
&&H(\boldsymbol{\phi},\mathbf{I})=\sum_i \omega_i I_i + \nonumber\\ &&\hspace*{1cm}-\frac{K}{N}\sum_{i,j} A_{ij} \sqrt{I_i I_j}(I_j-I_i)\rho(\phi_j-\phi_i)\, ,
\label{eq:Ham}
\end{eqnarray} 
where $\boldsymbol{\phi}=(\phi_1, \dots \phi_N)$ and $\textbf{I}=(I_1, \dots, I_N)$ are respectively the vectors of the angles variables and the actions variables. The previous model represents a class of systems able to describe the Lipkin-Meshkov-Glick (LMG) model in the thermodynamic limit \cite{lipkin} and of the Bose-Einstein condensate in a tilted optical lattice \cite{BEC}. The function $\rho$ describes the non-linear interaction among the pairs of oscillators while the parameter $K/N$ provides its strength. The adjacency matrix $A_{ij}$ encodes the connections among the oscillators. Observe that the classic Kuramoto model corresponds to $A_{ij}=A_{ji}=1$, $\forall i\neq j$, $A_{ii}=0$ and $\rho(x)=\sin x$. 

\begin{figure*}
\centering
\includegraphics[scale=0.415]{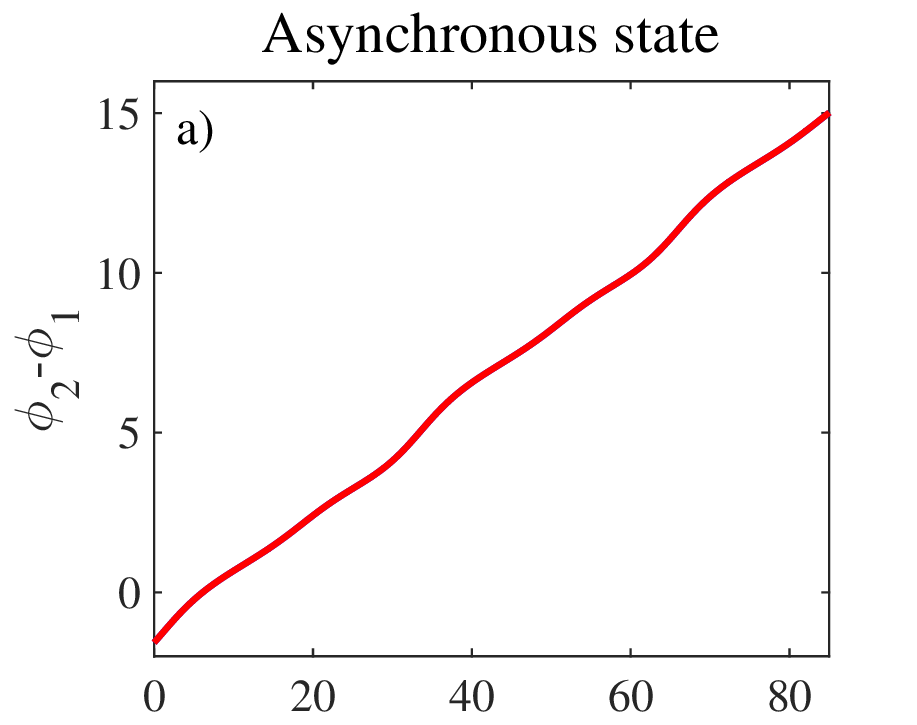}\hspace{-.45cm}
\includegraphics[scale=0.415]{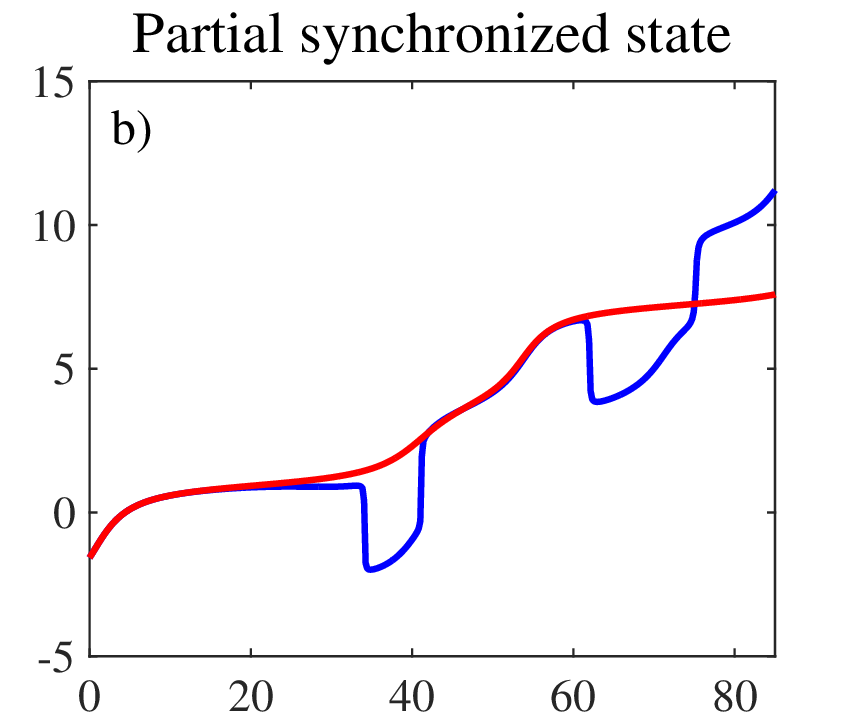}\hspace{-.45cm}
\includegraphics[scale=0.415]{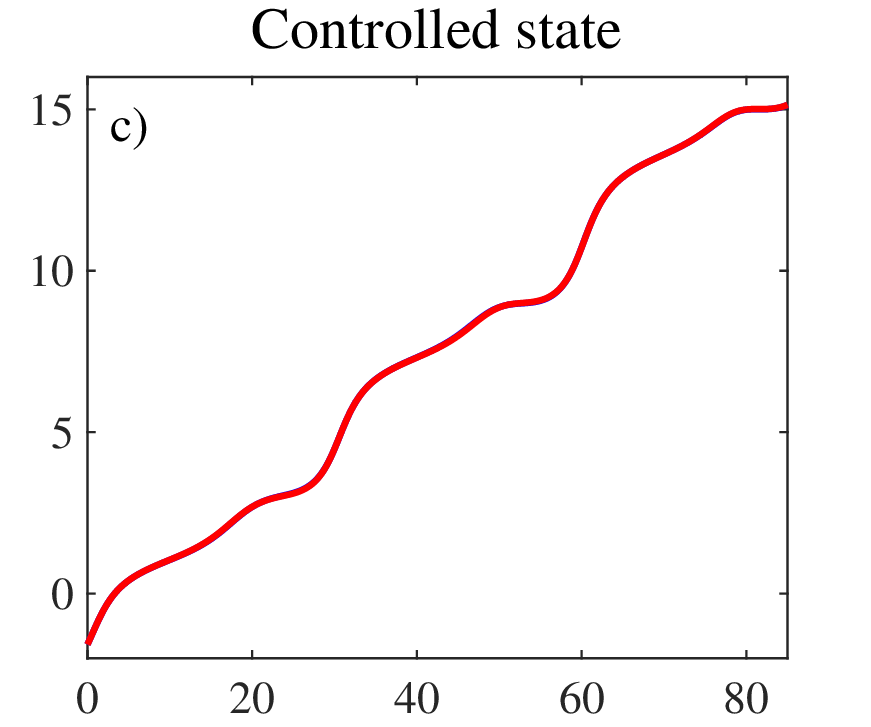}\\
\includegraphics[scale=0.415]{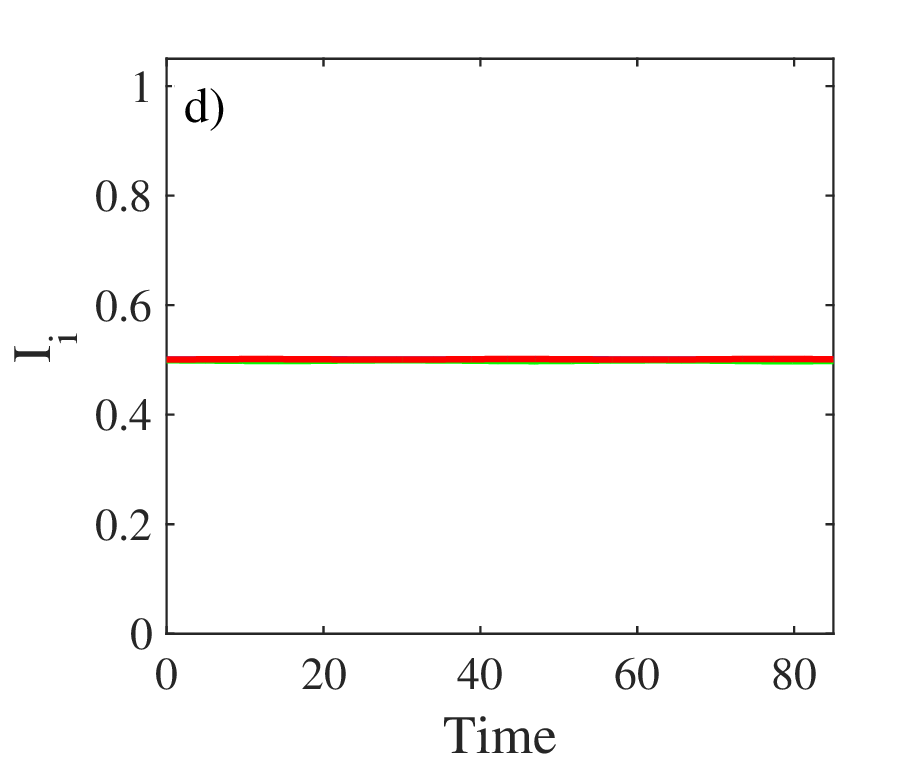}\hspace{-.6cm}
\includegraphics[scale=0.415]{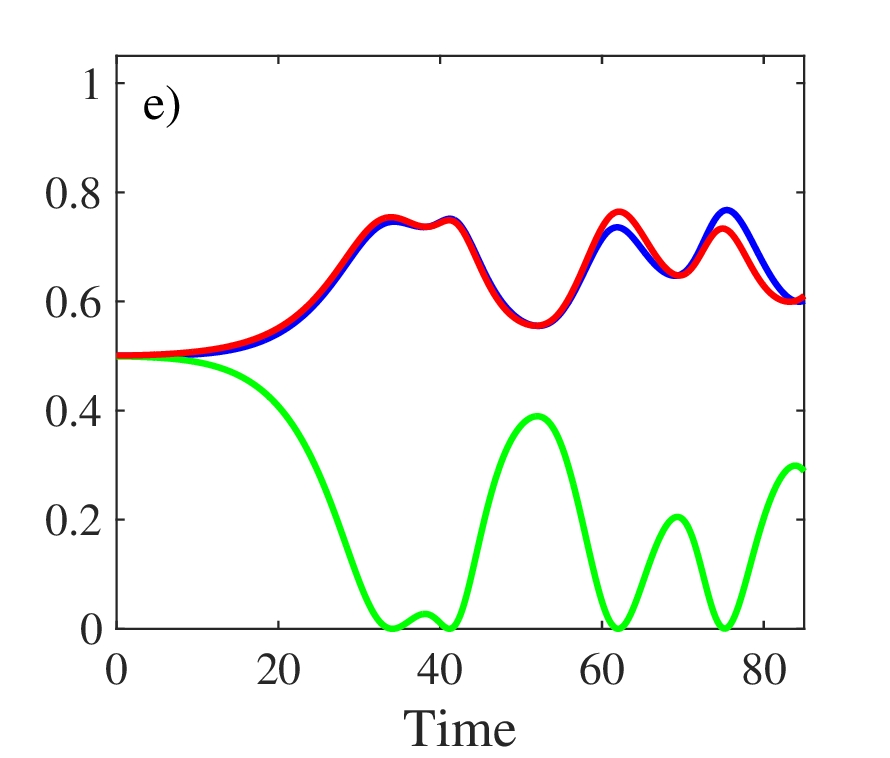}\hspace{-.6cm}
\includegraphics[scale=0.415]{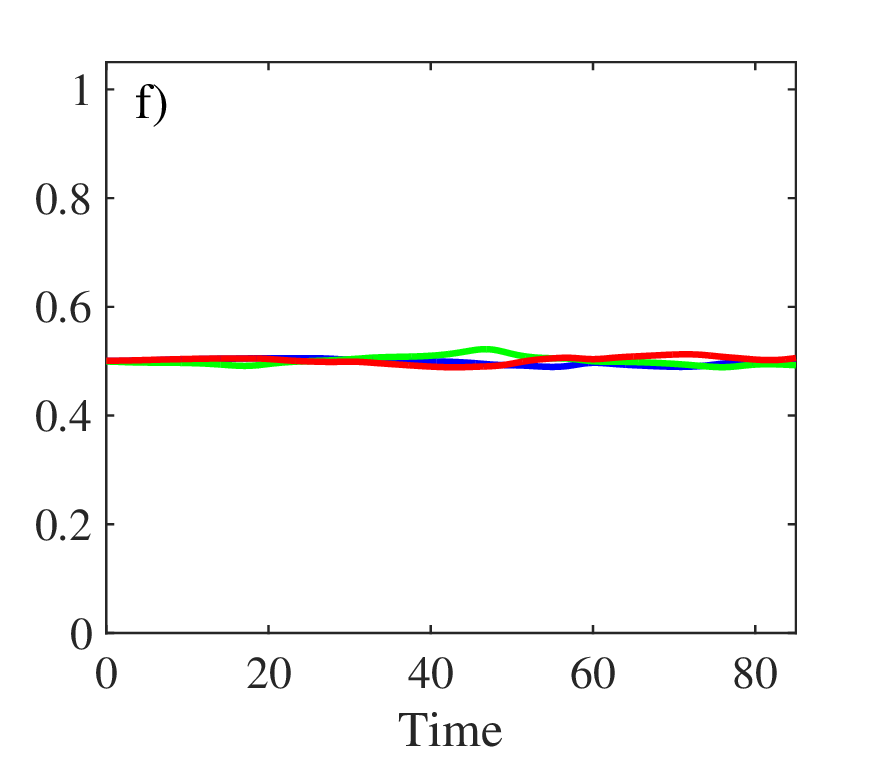}
\caption{Time evolution of action-angle variables for the Hamiltonian KM and of the original KM composed by three coupled oscillators. The angles drift once the actions remain close to the Kuramoto torus (panel $d$), as one can appreciate from panel $a$ where we report the phase difference between the first and the second oscillator for both the Hamiltonian KM and the original one, the curves are extremely close and thus one cannot differentiate between the two models. On the contrary partial phase-locked regime - horizontal plateaux - is present in panel $b$ (Hamiltonian model in blue and the original KM in red) corresponding to instability of the actions near the Kuramoto torus (panel $e$); let us observe that now the angle variables behave differently in the Hamiltonian model and in the original KM. Instead, if the KM is controlled the angles drift again (panel $c$) and the action turn stable (panel $f$). Once again the dynamics of the two systems are extremely close and thus one cannot differentiate between the two models. Initial conditions have been randomly chosen as a small perturbation of order $10^{-4}$ near the Kuramoto torus $I_i=1/2$, angle are uniformly random distributed in $[0,2\pi]$, the natural frequencies are set $(\omega_1,\omega_2,\omega_3) = (0.8,1, 1.2)$ and the coupling constant is set to $K=0.1$ in panels $a$, $d$ and to $K=0.3$ in panels  $b$, $c$, $e$, $f$.}
\label{fig:Fig1}
\end{figure*}

The time evolution of the angle-action variables is obtained from the Hamilton equations:
\begin{eqnarray}
\dot{I}_i &=& -\frac{\partial H}{\partial \phi_i}=-2\frac{K}{N}\sum_{j=1}^N A_{ij}\sqrt{I_iI_j}\left(I_j-I_i\right)\rho'(\phi_j-\phi_i)\nonumber\\ \label{eq:Idot}\\
\dot{\phi}_i &=& \frac{\partial H}{\partial I_i}=\omega_i + \frac{K}{N}\sum_{j=1}^N A_{ij}\bigg[2\sqrt{I_iI_j}\rho(\phi_j-\phi_i)+\nonumber\\&&\hspace*{2.5cm}-\sqrt{I_j/I_i}\left(I_j-I_i\right)\rho(\phi_j-\phi_i)\bigg]
\label{eq:phidot}
\end{eqnarray}
for $i=1,\dots,N$. One can readily realize that $I_i=J$ $\forall i$ are constants of motion for any fixed $J>0$, we can thus define the Kuramoto torus~\footnote{The dynamics of the Hamiltonian system~\eqref{eq:Ham} is invariant for the rescaling $I_i\rightarrow I_i/C$ and $K\rightarrow K/C$ where $C^2=2\sum_i I_i$ is a third first integral, besides the energy and $I_i=J$. One can thus set $C^2=N$ which corresponds to set $J=1/2$.}, $\mathcal{T}^{K}:=\{ (\mathbf{I},\boldsymbol{\phi})\in\mathbb{R}^N_+\times \mathbb{T}^N: I_i=1/2 \; \forall i\}$, moreover the restriction of Eq.~\eqref{eq:phidot} to the latter gives:
\begin{equation}
\dot{\phi}_i = \omega_i + \frac{K}{N}\sum_{j=1}^N A_{ij}\rho(\phi_j-\phi_i).
\label{eq:kurmod}
\end{equation}
It has been proved~\cite{timme} that when the Kuramoto oscillators of (\ref{eq:kurmod}) enter in a synchronisation phase, then the dynamics of the actions close to the Kuramoto torus, become unstable and exhibit a chaotic behaviour (see Fig.~\ref{fig:Fig1}). 
Based on the above remark, our aim is to eliminate (reduce) the synchronisation from the KM~\eqref{eq:kurmod} by controlling the Hamiltonian system~\eqref{eq:Ham} by adding a \lq\lq small\rq\rq control term able to increase the stability of the invariant torus $\mathcal{T}^{K}$ and thus to impede the phase-lock of the coupled oscillators. Let us observe that this approach is completely different from the classical feedback control method where one adds a term, usually of the same order of magnitude of the uncontrolled system, depending on some measurement of the system state, that can also produce a time-delay in the dynamics.

To be more precise let us rewrite the Hamiltonian~\eqref{eq:Ham} in the form $H=H_0+V$, where $H_0$ is the integrable part, i.e. the uncoupled harmonic oscillators, and $V$ the non-linear term that can be considered as a perturbation of $H_0$ because of the smallness of $K$. The main idea of Vittot and coworkers~\cite{vittot,ciraolo} is to add a small control term $f(V)\sim \mathcal{O}(V^2)$ to $H$ in order to reduce the impact of the perturbation $V$, roughly speaking to increase the stability of the invariant torus, eventually obtaining the controlled Hamiltonian function:
\begin{equation}
H^{ctrl} \equiv H_0 + V + f(V)\, .
\label{eq:Hcontrol}
\end{equation}
In terms of costs the size of $f(V)$ implies that the controlling procedure is much less invasive that other techniques generally used in control theory and also able to give a prompt response to possible abnormal dynamics without time lags. 

Starting from $H_0$ one can construct a linear operator $\{H_0\}$ defined on the vector space of $\mathcal{C}^{\infty}$-real functions defined on the phase space, such that $\{H_0\}f:=\{H_0,f\}$ is the Poisson bracket between $H_0$ and $f$. Then the control term can be explicitly obtained~\cite{vittot} as
\begin{equation}
f(V) = \sum_{n \geq 1} \frac{\{-\Gamma V \}^n}{(n+1)!}(n \mathcal{R} V + V)= f_1+f_2+\dots \, ,
\label{eq:fVserie}
\end{equation}  
where $\Gamma$ is the pseudo-inverse operator of the Hamiltonian function $H_0$ and $\mathcal{R} V$ the resonant part of the perturbation term $V$~\footnote{More information concerning the Hamiltonian control are provided in the Supplementary Material. We also invite the interested reader to consult~\cite{vittot} to have more details about this formula and the reason why one should work with a pseudo-inverse operator instead of the inverse one.}. For a sake of simplicity and without loss of generality we can assume $\boldsymbol{\omega}=(\omega_1,\dots,\omega_N)$ to be not resonant, i.e. for all $\textbf{k}\in\mathbb{Z}\setminus\{0\}$ then $\textbf{k}\cdot \boldsymbol{\omega} \neq 0$, in this case Eq.~\eqref{eq:fVserie} simplifies because $\mathcal{R} V=0$.

The embedding of the KM into the Hamiltonian~\eqref{eq:Ham} and thus the possibility to control the former by controlling the latter, is based on the existence of the invariant torus $\mathcal{T}^{K}$. However the controlled Hamiltonian $H^{ctrl}$ no longer verifies such property once one uses the full control term $f(V)$, one can nevertheless provide an {\em effective} control by truncating the latter to it first term $f_1\sim \mathcal{O}(V^2)$. Because $H^{ctrl}_1=H_0+V+f_1$ preserves the Kuramoto torus (for further details see the Supplementary Material) one can transfer this information into the KM and thus determine an effective control term:
\begin{equation}
\dot{\phi}_i=\omega_i + \frac{K}{N}\sum_{j=1}^N A_{ij}\rho(\phi_j-\phi_i)+\dot{\phi}^{ctrl}_i\, ,
\end{equation}
with
\begin{widetext}
\vspace*{-.55cm}
\begin{align}
\dot{\phi}^{ctrl}_i = &-\frac{K^2}{4 N^2}\left[\sum_j A_{ij}\rho'(\phi_j-\phi_i)\sum_l\frac{A_{il}}{\omega_l-\omega_i}\rho'(\phi_l-\phi_i)+\sum_j\frac{A_{ij}}{\omega_j-\omega_i}\rho(\phi_j-\phi_i)\sum_l A_{il}\rho(\phi_l-\phi_i)+\right.\nonumber\\&\left.-\sum_l\left(A_{il}\rho'(\phi_i-\phi_l)\sum_j\frac{A_{jl}}{\omega_j-\omega_l}\rho'(\phi_j-\phi_l)+\frac{A_{il}}{\omega_i-\omega_l}\rho(\phi_i-\phi_l)\sum_j A_{jl}\rho(\phi_j-\phi_l)\right)\right].
\label{eq:kurcontrol}
\end{align}
\vspace*{-.2cm}
\end{widetext}
where with a slight abuse we used the same letter to denote the new angular variable. 
%\begin{widetext}
%\vspace*{-.55cm}
%\begin{align}
%\dot{\phi}^{ctrl}_i = &-\frac{I^2 K^2}{2 N^2}\left[\sum_i(A_{ji}+A_{ij})\rho'(\phi_i-\phi_j)\sum_i\frac{A_{ji}+A_{ij}}{\omega_i-\omega_j}\gamma'(\phi_i-\phi_j)+\sum_i\frac{A_{ji}+A_{ij}}{\omega_i-\omega_j}\gamma(\phi_i-\phi_j)\sum_i(A_{ij}+A_{ij})\gamma(\phi_i-\phi_j)+\right.\nonumber\\&\left.-\sum_l\left((A_{lj}+A_{jl})\gamma'(\phi_j-\phi_l)\sum_i\frac{A_{li}+A_{il}}{\omega_i-\omega_l}\gamma'(\phi_i-\phi_l)+\frac{A_{lj}+A_{jl}}{\omega_j-\omega_l}\gamma(\phi_j-\phi_l)\sum_i(A_{li}+A_{il})\gamma(\phi_i-\phi_l)\right)\right].
%\label{eq:kurcontrol}
%\end{align}
%\vspace*{-.2cm}
%\end{widetext}

In Fig.~\ref{fig:Fig1} (panels $c$, $f$) we report the results of some numerical simulations of the controlled KM~\eqref{eq:kurcontrol}; one can clearly appreciate that even for relatively large coupling  $K=0.3$, the angles are drifting and the actions become stable again while in the uncontrolled KM they where phase-locked (see panels $a$, $d$ and $b$, $e$ in Fig.~\ref{fig:Fig1}).
%\begin{figure}[h!]
%\vspace*{-.25cm}\hspace*{-.7cm}
%\includegraphics[width=1.2\columnwidth]{Fase_control.eps}\\
%\hspace*{-.7cm}
%\includegraphics[width=1.2\columnwidth]{Azioni_control.eps}
%\caption{The Hamiltonian control of the synchronization: the panels show the dynamics of three coupled oscillators in ($a$, $b$) synchronous and ($c$, $d$) controlled (asynchronous) regime respectively. The parameters are the same as in Figure \ref{fig:Fig1}. (NON CAPISCO UNA COSA: SE USI f1 IN H PERCHE IL TORO I=1/2 NON E' ESATTAMENTE CONSERVATO? E' PER VIA DELLE CONDIZIONI INIZIALI CHE NON SONO ESATTAMENTE 1/2? I PANNELI a) E b) SONO GLI STESSI DI FIG1, NON CONVEREBBE METTERE TUTTO INSIEME?}
%\label{fig:Fig2}
%\end{figure}

\begin{figure}
\hspace*{-.4cm}
\includegraphics[width=1.1\columnwidth]{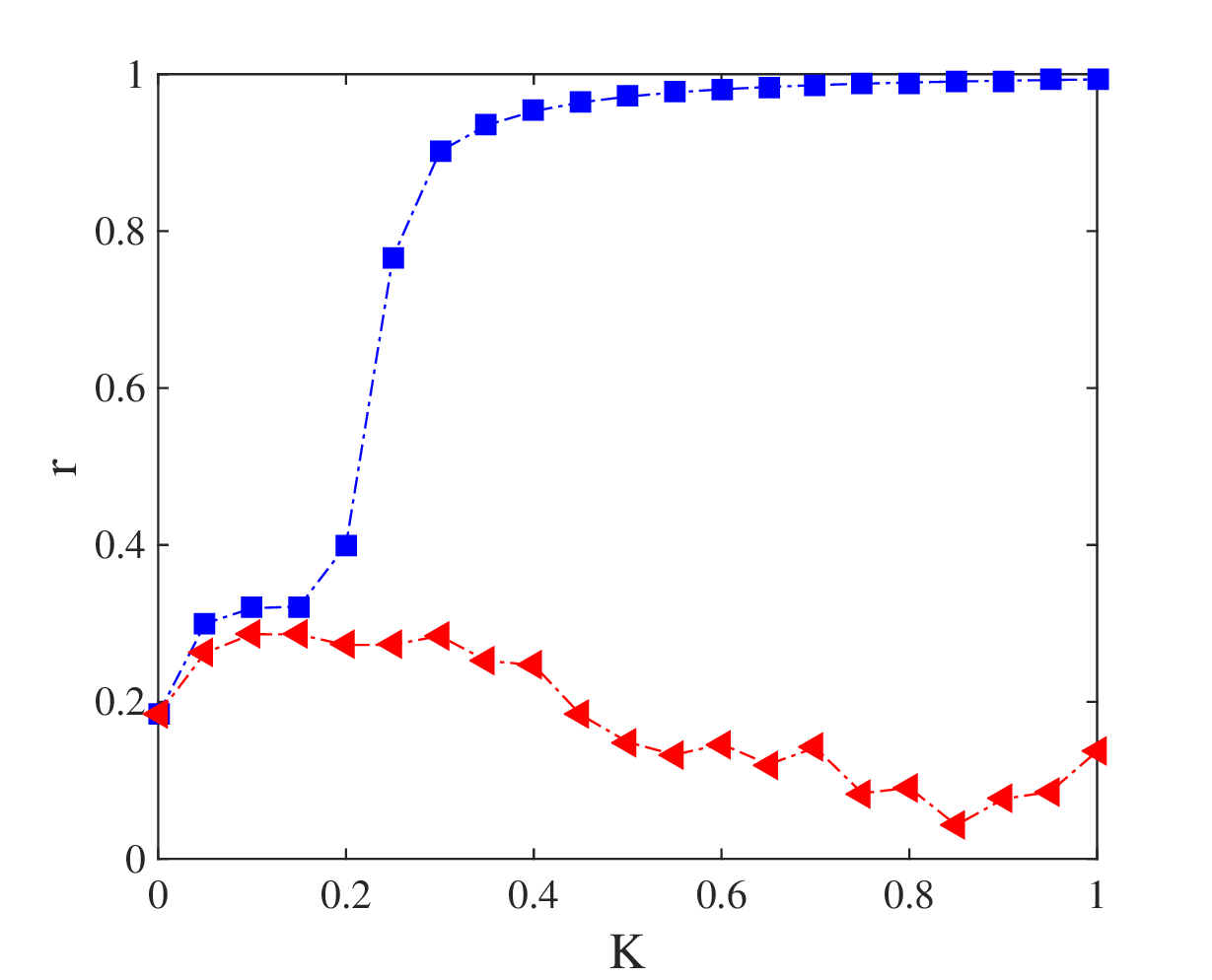}
\caption{The order parameter $r$ as a function of the coupling coefficient $K$ for $20$ coupled oscillators averaging over a time interval of $250$ unities. In the controlled model (red triangles) the coupled oscillators are prevented from synchronisation even for ranges of $K$ where the original model (blue squares) do synchronise. The remaining parameters are set as in Fig.~\ref{fig:Fig1}.}
\label{fig:Fig3}
\end{figure}

A macroscopic index is often used to measure the strength of the synchronization, the order parameter as originally proposed by Kuramoto~\cite{kuramoto}
\begin{equation}
re^{i\psi} = \frac{1}{N} \sum_{j=1}^N e^{i\phi_j}\, .
\label{eq:par_ord}
\end{equation}
If $r\sim 0$, the oscillators are almost independent each other while if $r\sim 1$ they are close to phase-lock. In Fig.~\ref{fig:Fig3} we present the results of numerical simulations of the original KM and of the controlled one both involving $20$ oscillators, by showing the behaviour of the order parameter as a function of the coupling strength $K$. It is a well known result that there exists a synchronisation threshold $K_c$ above which $r\rightarrow 1$ and the oscillators of the KM do synchronise (blue curve in Fig.~\ref{fig:Fig3}), let us however observe that in the controlled model, the oscillators remain well independent for $K$ much larger than $K_c$. This definitively confirms the goodness of the proposed effective control. As a side remark, let us underline that the results hold for the general case for $N$ coupled oscillators. 

In conclusion, in this work we have developed a new method to control the synchronisation of nonlinearly coupled oscillators based on the Hamiltonian control formalism, the idea has been hereby applied to the paradigmatic Kuramoto model but it is far more general. In this letter we have introduced a controlled system aimed to prevent the phase-locking, i.e. adapted to systems where the synchronisation can induce undesired negative effects on the global dynamics as was the case of the now famous Millennium Bridge of London. Nevertheless with our method we can also deal with the opposite case where we want to control the system to enhance the synchronisation, in this case a resonant control will do the required job. We are thus confident that our approach can be a successful candidate to afford the problem of the control of general synchronisation processes.

\appendix*

\section*{Supplementary Material\\\vspace*{.2cm}\small{Hamiltonian control theory}}

The results presented in the main text are developed in the framework of the Hamiltonian control theory. The aim of this section is to provide a short introduction to the subject and to add some details of the computations presented in the main text. The starting point is to consider an Hamiltonian system written as the sum of two parts, the integrable Hamiltonian $H_0$ and the perturbation term $V$ of order $\mathcal{O}(\epsilon)$ (where $\epsilon$ is a small parameter)
\begin{equation}
 H = H_0 + V\, .
 \label{eq:AP_Ham}
\end{equation}

The goal of the Hamiltonian control theory is to slightly modify  eq. (\ref{eq:AP_Ham}) by adding a control term, small in front of $V$, in such a way the new controlled hamiltonian $H^{ctrl} = H_0 + V+f$ exhibits some suitable features, for instance to improve the regularity of the dynamics, because we expect a much lesser chaotic dynamics and the appearance of regular structures such as invariant tori in the phase plane. 

To achieve this goal we should assume the existence of an operator $\Gamma: \mathcal{A}\rightarrow \mathcal{A}$, where $\mathcal{A}$ is the Lie algebra of the functions defined on the phase space, such that verifies 
\begin{equation}
\label{eq:pseudoinv}
\{ H_0 \}^2 \Gamma = \{ H_0 \}.
\end{equation}
where $\{\cdot\}$ are the Poisson brackets and do satisfy $\{V\}W=-\{W\}V$ (antisymmetry) and $\{\{V\}W\}=\{V\}\{W\}-\{W\}\{V\}$ (Jacobi identity), for all $V,W \in\mathcal{A}$. We call such operator the \textit{pseudo-inverse} of $\{ H_0 \}$, observe that  in general it is not unique; because $\{H_0\}$ has a non trivial kernel (e.g. $\{H_0\} H_0 = 0$), Eq.~\eqref{eq:pseudoinv} is the minimal requirement to have a sort of inversion. 

From $ \Gamma$ we define two other operators: the \textit{non-resonant} operator $\mathcal{N}$ and the \textit{resonant} one $\mathcal{R}$ as follows:
\begin{equation*}
\mathcal{N} = \{ H_0 \} \Gamma
\end{equation*}
\begin{equation*}
\mathcal{R} = \mathds{1} - \mathcal{N} 
\end{equation*}
where $\mathds{1}$ is the identity of the Lie algebra $\mathcal{A}$. 

Let us denote by $e^{t\{H\}}$ the flow generated by $H$, that is $e^{t\{H\}}=\sum_{n\geq 0}t^n\{H\}^n/n!$, then the main result of~\cite{vittot} can be stated as follows, under the previous assumptions the following conjugation formula does hold
\begin{equation*}
\forall t \in \mathbb{R},\;\; e^{t\{H_0 + V + f(V) \}} = e^{-\{ \Gamma V\}} e^{t\{H_0\}} e^{t \{ \mathcal{R} V \}} e^{\{\Gamma V\}} 
\end{equation*}
where the function $f:\mathcal{A} \rightarrow \mathcal{A}$ is defined by
\begin{equation}
\label{eq:fform}
f(V) = \sum_{k \geq 1} \frac{(-1)^k\{\Gamma V \}^k}{(k+1)!}(k \mathcal{R} V + V)\, .
\end{equation}

The previous formula means that if we add the small control term $f(V)$ of order $\mathcal{O}(V^2)$ to $H_0+V$, then the orbits of the controlled version of the system (\ref{eq:AP_Ham}) will coincide with the ones of the unperturbed system $H_0$ except for the (possible) presence of the resonant term $\mathcal{R}V$. This makes the controlled orbits much more regular then the ones of the original uncontrolled system.  

Let us observe that if the Hamiltonian system $H_0$ can be written in action--angle variables, $(\mathbf{I},\boldsymbol{\phi})$, in the vicinity of an elliptic equilibrium, that is $H_0=\boldsymbol{\omega}\cdot \mathbf{I}$, for some frequency vector $\boldsymbol{\omega}\in\mathbb{R}^n$, then the pseudo-inverse operator can be formally written as:
\begin{equation}
\label{eq:pseudoinv2}
\Gamma = \frac{1}{\boldsymbol{\omega}\cdot \partial_{\boldsymbol{\phi}}}\mathcal{N}\, ,
\end{equation}
If the frequency vector is non-resonant, namely $\boldsymbol{\omega \cdot k} \neq 0$ for all $\boldsymbol{k} \in \mathbb{Z}^n \backslash {0}$, then $\mathcal{R}V$ reduces to the constant term of $V(\mathbf{I},\boldsymbol{\phi})$ once written in Fourier series.

Let us conclude by applying the previous theory to the Hamiltonian model presented in the main text Eq. (2) containing as particular case to the Kuramoto model:
\begin{eqnarray*}
H(\boldsymbol{\phi},\mathbf{I})&=&\sum_i \omega_i I_i -\frac{K}{N}\sum_{i,j} A_{ij} \sqrt{I_i I_j}(I_j-I_i)\rho(\phi_j-\phi_i)\\&\equiv& H_0(\mathbf{I})+V(\boldsymbol{\phi},\mathbf{I})\, ,
\end{eqnarray*} 
where $H_0$ and $V$ are defined by the rightmost equality.

Under the assumption of non-resonant frequency we can write the action of the operator $\Gamma$ on the perturbation term $V$ as follows:
\begin{equation*}
\Gamma V = \frac{K}{N}\sum_{\substack{i,j\\i\neq j}}\frac{A_{ij}}{\omega_i-\omega_j}\sqrt{I_i I_j}(I_i-I_j)\rho'(\phi_i-\phi_j)\, .
\end{equation*}

Using the explicit form of $V$ one straightforwardly get
\begin{equation*}
\mathcal{R}V = 0 \;\; \text{and}\;\; \mathcal{N}V = -\frac{K}{N}\sum_{i,j} A_{ij}\sqrt{I_i I_j}(I_j-I_i)\rho(\phi_j-\phi_i)=V.
\end{equation*}
In this case the control term $f$ given by Eq.~\eqref{eq:fform} can be calculated through a recursive procedure:
\begin{eqnarray*}
&f = \sum_{s\geqslant2} f_s \quad \mathrm{where} \quad f_1=V\\&\text{and for all $s\geq 2$}\quad
f_s=-\frac{1}{s}\lbrace \Gamma V, f_{s-1}\rbrace\sim \mathcal{O}(K^s/N^s)\, ,
\end{eqnarray*}   
once we set $V\sim \mathcal{O}(K/N)$.

To obtain the control term for the Kuramoto model embedded in the Hamiltonian formalism is now matter of computation starting form the previous equation.

Let us finally remark that the previous results can be generalised to the case of resonant frequencies as well~\cite{carletti}.  
  
\end{document}